\documentclass[preprint,aps,prb]{revtex4}

\usepackage{epsf}
\usepackage{bm}

\begin{document}
    
\title{Thermodynamics of tin clusters}

\author{Kavita Joshi and D. G. Kanhere}

\affiliation{%
Department of Physics, University of Pune, 
Ganeshkhind, Pune 411 007, India}

\author{S. A. Blundell}

\affiliation{%
D\'epartement de Recherche Fondamentale sur la 
Mati\`ere Condens\'ee, CEA-Grenoble/DSM \\
17 rue des Martyrs, F-38054 Grenoble Cedex 9, France}

\date{\today}

\begin{abstract} 
We report the results of detailed thermodynamic investigations of the
Sn$_{20}$ cluster using density-functional molecular dynamics.  These
simulations have been performed over a temperature range of 150 to
3000~K, with a total simulation time of order 1~ns.  The prolate
ground state and low-lying isomers consist of two tricapped trigonal
prism (TTP) units stacked end to end. The ionic specific heat,
calculated via a multihistogram fit, shows a small peak around 500~K
and a shoulder around 850~K. The main peak occurs around 1200~K, about
700~K higher than the bulk melting temperature, but significantly
lower than that for Sn$_{10}$.  The main peak is accompanied by a
sharp change in the prolate shape of the cluster due to the fusion of
the two TTP units to form a compact, near spherical structure with a
diffusive liquidlike ionic motion.  The small peak at 500~K is
associated with rearrangement processes within the TTP units, while
the shoulder at 850~K corresponds to distortion of at least one TTP
unit, preserving the overall prolate shape of the cluster.  At all
temperatures observed, the bonding remains covalent.
\end{abstract}

\pacs{31.15.Qg,36.40.Sx,36.40.Ei,82.20.Wt}

\maketitle

%
%
\section{Introduction\label{sec:intro}}

Clusters of the semiconducting Group IV elements are interesting not
only for their potential technological applications, but also from a
fundamental point of view.  Small clusters of Si, Ge, and Sn have an
unusual size-dependent structural evolution.  Ion mobility
measurements\cite{Simob,JarBow92,Gemob,Alex1} revealed that for small
sizes the clusters are prolate, with aspect ratios up to three, but
undergo a rearrangement to a more compact, spherical form at a size of
a few tens.  The rearrangement is rather abrupt for Si and Ge, but
more gradual for Sn.  Density-functional theory (DFT) studies of
ground-state
structures\cite{Jack96,VasOgut97,HoShvart98,LiuLu98,Stott98,%
ShvartLiu99,Kum}
and optical response\cite{SiTTP} confirmed the prolate shape of the
small sizes and showed further that many of the ground-state
structures in this size range contain the tricapped trigonal prism
(TTP) unit, a nine-atom unit consisting of a triangular prism capped
by one atom on each of the three side faces.

Recently, interest was focused on the thermodynamics of the prolate Sn
clusters after Shvartsburg and Jarrold\cite{Alex} found evidence,
based on measurements of ionic mobility, that these clusters remain
solidlike at temperatures up to at least 50~K {\em higher} than the
melting point of bulk Sn [$T_{m}(\mathrm{bulk}) = 505$~K].  This is
the only example so far observed of a small particle melting at a
higher temperature than the bulk.  It contradicts the standard
paradigm, based on thermodynamic arguments, that a cluster should melt
at a lower temperature than the bulk because of the effect of the
cluster surface.\cite{pawlow,mesotherm,Buff76} Many examples of a
melting-point depression for clusters of mesoscopic size (greater than
a few thousand atoms) have been demonstrated
experimentally,\cite{Buff76,mesoexp} and recently a depression was 
also observed for small Na clusters in the size range of 50 to
200.\cite{Freiburg}

There have been a few DFT studies that have implications for the
thermodynamics of Sn clusters.  Lu {\em et al.}\cite{Lu} carried out
simulated-annealing studies of Sn$_{N}$ clusters ($N \leq 13$), within
the local-density approximation (LDA), in which the clusters were
cooled from a random atomic configuration at 1980~K down to 0~K in
22~ps.  By following the root-mean-square atomic displacements as a
function of time, they found that the final ground-state geometries
were reached roughly half way through the annealing process, at about
1000~K, for $10 \leq N \leq 13$.  Majumder {\em et al.}\cite{Kum}
calculated energies of some low-lying isomeric structures within the
generalized gradient approximation (GGA) for the exchange-correlation
functional, and pointed out that the binding energy per atom was
unusually high, only about 11\% less than that of the bulk.  Finally,
we studied the thermodynamics of the Sn$_{10}$
cluster,\cite{sn10} the smallest Sn cluster whose ground state
displays the TTP unit.  Using DFT in the LDA, we found the TTP unit to
break up for temperatures of about 1600~K or higher, yielding a broad
peak in the canonical specific heat around 2300~K.

In this paper, we revisit the thermodynamics of Sn clusters in greater
detail.  Prolate structures are identified in ion mobility experiments
by having a smaller mobility than a spherical structure.  Shvartsburg
and Jarrold\cite{Alex} observed that the mobility of Sn$_{N}^{+}$
clusters ($10 \lesssim N \lesssim 30$) remained small even at 50~K
above the bulk melting point.  They argued that, if the clusters had
melted, they would have collapsed to a more compact, spherical form,
observable by an increase in mobility, contrary to the actual
observation.  Now, the cluster Sn$_{10}$, which we studied
earlier,\cite{sn10} contains just one TTP unit and already has a
compact structure both in its ground state and in high-temperature
liquidlike states.  Thus one would not expect to be able to detect its
melting in a mobility experiment.  We have therefore extended our
methods to treat a cluster of twice the size, Sn$_{20}$.  In its
ground state, Sn$_{20}$ has a prolate structure, with an aspect ratio
of about 2.5, consisting of two TTP units stacked end to
end. In this paper we study the structural rearrangements,
caloric curve, and electronic properties of this prolate cluster,
and contrast the results with those of Sn$_{10}$.

As in our earlier paper,\cite{sn10} we use {\em ab initio}
Born-Oppenheimer molecular dynamics, within the LDA of DFT, and
norm-conserving pseudopotentials.  This first-principles treatment of
electronic structure is highly desirable, since the directional
covalent bonding in small Sn clusters would be hard to describe
reliably by a parametrized potential, simultaneously for the ground
state and for the higher-lying isomers and liquidlike forms crucial
for understanding the melting properties.  Despite the computational
expense of {\em ab initio} approaches, we obtain a sufficiently long
statistical sampling of the ionic phase space as to converge the
canonical specific heat to of order 10\%, as extracted from a
multihistogram fit.  Our {\em total\/} sampling time is of order 1~ns. 
We believe that this statistical convergence criterion on the specific
heat gives us a good sampling of the various dynamical and rearrangement
processes of the cluster as a function of temperature.

As we shall see, the Sn$_{20}$ cluster undergoes a number of dynamical
processes as a function of temperature.  One of these is a collapse of
the prolate structure toward a more spherical, disordered structure. 
This collapse, which is the point where an increase in ionic
mobility would be observed, occurs about 700~K above the bulk
melting point in our LDA model.  Thus our calculations confirm, and
explain, the experimentally observed stability of the prolate form at
50~K above the bulk melting point.\cite{Alex} However, at lower
temperatures, of order the bulk melting point or lower, we also find a
large number of dynamical processes that lead to a permutational
rearrangement of the atoms of the cluster, but which preserve the
overall prolate shape and would therefore be unobservable in a
mobility measurement.

Since we perform DFT calculations of electronic structure for every
ionic configuration in our thermal ensembles, we can also study
electronic properties as a function of temperature.  Noting that solid
bulk Sn undergoes a phase transition at 286~K between a
low-temperature covalent form and a higher-temperature metallic form,
we examine the bonding characteristics of the various isomers of Sn$_{20}$
using the electron localization function\cite{Silvi} (ELF). 
The bonding is found to be predominantly covalent at all temperatures
considered.  However, the gap between the highest occupied molecular
orbital and the lowest unoccupied molecular orbital (HOMO-LUMO gap)
is reduced dramatically during the collapse of the prolate structure.
Finally, although
we study only Sn clusters in this paper, we expect that a large number
of our results would be qualitatively similar for Si and Ge
clusters, since the structure and bonding of these clusters is 
similar to those of Sn.

The paper is organized as follows.  In Sec.~\ref{sec:comput} we
describe briefly our computational and statistical approaches.  
Results for the structural, caloric, and
electronic properties are given in Sec.~\ref{sec:results},
and the conclusions are given in Sec.~\ref{sec:conclusions}.

%
%
\section{Computational Details\label{sec:comput}}   

All the calculations have been performed using the Kohn-Sham
formulation of density-functional molecular dynamics (MD),\cite{Payne}
using the nonlocal norm-conserving pseudopotentials of Bachelet {\em
et al.}\ \cite{BHS} in Kleinman-Bylander form \cite{KB} with the
Ceperley-Alder \cite{CA} exchange-correlation functional.  We use a cubic
supercell of length 40~a.u.\ with an energy cutoff of 16.3~Ry, which is found
to provide a convergence of the total electronic energy of better than
10$^{-6}$~a.u. To have a sufficiently precise evaluation of the
Hellmann-Feynman forces on the ions, we ensure that the residual norm
of each KS orbital, defined as $|H_{{\rm KS}}\psi _{i}-\epsilon
_{i}\psi _{i}|^{2}$ ($\epsilon _{i}$ being the eigenvalue
corresponding to eigenstate $\psi _{i}$ of the KS Hamiltonian $H_{{\rm
KS}}$), was maintained at $10^{-9}$~a.u.  The ground state and other
equilibrium structures have been found by carrying out several 
steepest-descent runs starting from structures chosen periodically 
from a high-temperature simulation.  

The ionic phase space is sampled by isokinetic MD,\cite{isokin} 
in which the kinetic energy is held constant using velocity scaling. We
split the temperature range 150 K $\leq T\leq 3000$~K into about 30
different constant temperatures, performing up to 40~ps of simulation
for each.  We observe no evaporation on the time scale of our
simulations, and therefore a reflective container (often used with
classical potentials) is not explicitly required. Following the
multiple-histogram (MH) method,\cite{MH} we then extract the classical 
ionic density of states $\Omega (E)$ of the system, or equivalently the 
classical ionic entropy $S(E)=k_{B}\ln
\Omega (E)$.  With $S(E)$ in hand, one can evaluate
thermodynamic averages in a variety of ensembles.  We focus in this
work on the ionic specific heat and the caloric curve.  In the
canonical ensemble, the specific heat is defined as usual by
$C(T)=\partial U(T)/\partial T$, where $U(T)=\int E\,p(E,T)\,dE$ is
the average total energy, and where the probability of observing an
energy $E$ at a temperature $T$ is given by the Gibbs distribution
$p(E,T)=\Omega (E)\exp (-E/k_{B}T)/Z(T)$, with $Z(T) $ the normalizing
canonical partition function.  In the microcanonical ensemble, the
temperature $T(E)$ at energy $E$ is defined following the
thermodynamic definition as $T(E)=[\partial S(E)/\partial E]^{-1}$. 
The microcanonical specific heat is then defined as 
$C(E)=[\partial T(E)/\partial E]^{-1}$.
More details of the extraction of thermodynamic averages can be found 
in our earlier work.\cite{nathermo}

The nature of the bonding can be investigated using the
ELF\cite{Silvi} along with the charge density.  Such ELF have been
found to be extremely useful for elucidating the bonding
characteristics of a variety of systems, especially in conjunction
with the charge density.  For a single determinantal wavefunction
built from KS orbitals $\psi _{i}$, the ELF is defined as
\begin{equation}
\chi _{{\rm ELF}}=[1+{(D/D}_{h}{)}^{2}]^{-1}, 
\end{equation}
where
\begin{eqnarray} D_{h}&=&(3/10){(3{\pi
}^{2})}^{5/3}{\rho }^{5/3}, \\ D&=&(1/2)\sum_{i}{\
|{\bm{\nabla} \psi _{i}}|}^{2}-(1/8){|{\bm{\nabla}
\rho }|}^{2}/\rho, \end{eqnarray} with $\rho \equiv \rho
({\bf r})$ the valence-electron density.  A value of
$\chi _{{\rm ELF}}$ near 1 represents a perfect
localization of the valence electron
density.\cite{Silvi}

\begin{figure}
\epsfxsize=7.5cm
\centerline{\epsfbox{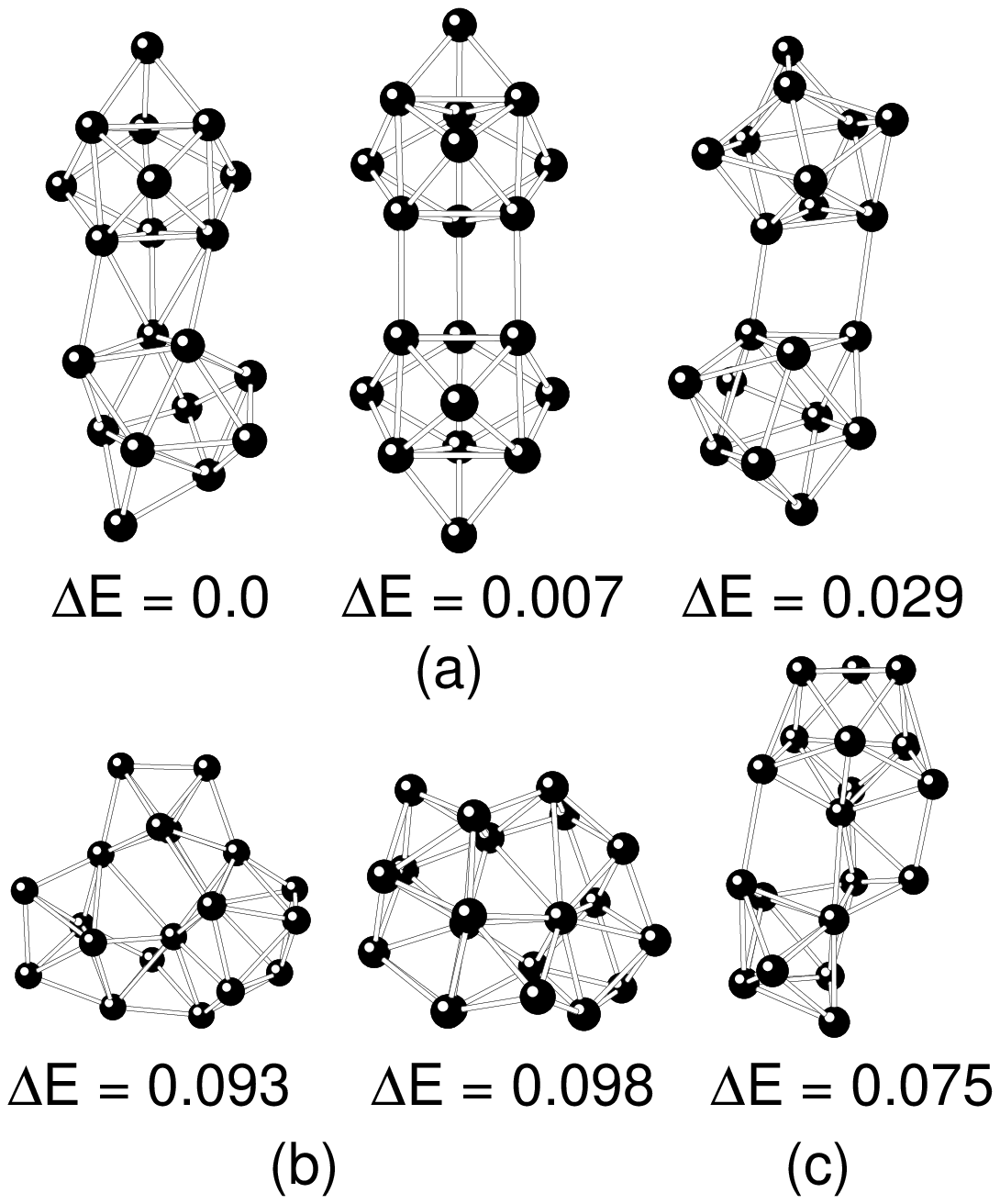}}
\caption{(a) Some prolate equilibrium 
geometries having two distinct TTP units.  The structure on the left
is the lowest-energy structure found. (b) Nonprolate equilibrium 
structures found at high temperature $T > 1250$~K. (c) Structure 
observed during the transition from prolate to nonprolate around 
$T = 1250$~K. The excitation energy relative to the ground state 
is shown as $\Delta E$, in units of eV per atom.\label{fig1}}
\end{figure}                      

%
%
\section{Results and Discussions\label{sec:results}}

We begin our discussion by presenting some equilibrium geometries of
Sn$_{20}$ (shown in Fig.\ \ref{fig1}).  We distinguish two broad
classes of equilibrium structures: one [Fig.\ \ref{fig1}(a)] where all
the structures have a prolate shape and show the presence of two
distinct TTP units, but differ in the relative orientation of the TTP
units, and a second one [Fig.\ \ref{fig1}(b)] with nonprolate, compact
geometries indicating that the TTP units have lost their identity.  In
Fig.~\ref{fig1}(c), we also show a typical intermediate structure
in which only one of the TTP units has lost its identity, but which is
still prolate.  
The most symmetric structure (proposed as the ground state in
Ref.~\onlinecite{Kum}) is slightly higher in energy than  our lowest-energy
structure. The only difference in these two structures is the relative
orientation of the TTP units. 
Evidently, changes in the relative orientation of the TTP
units give rise to several low-lying isomers
[two of which are shown in Fig.\ \ref{fig1}(a)] having
excitation energies of order 0.01~eV per atom relative to the ground
state.  It is also interesting to note that a recent study\cite{Si20} 
has found the most stable structure of Si$_{20}$ to be very similar to
that for Sn$_{20}$ found in the present work.

In Fig.\ \ref{fig1}(b), we have shown two nonprolate equilibrium
geometries having a difference in binding energy with respect to the
ground state of about 0.1~eV per atom.  These nonprolate structures
are somewhat disordered, although some basic units such as tetrahedra,
trigonal prisms, pentagonal bipyramids, etc., may be observed.  The
equilibrium structure shown in Fig.\ \ref{fig1}(c) is an intermediate
structure observed during the transition from prolate to nonprolate. 
In this geometry, one TTP unit is intact while the other is
completely destroyed.  This structure is 0.075~eV higher in
binding energy per atom compared to the ground state.

\begin{figure}
\epsfxsize=6.0cm
\centerline{\epsfbox{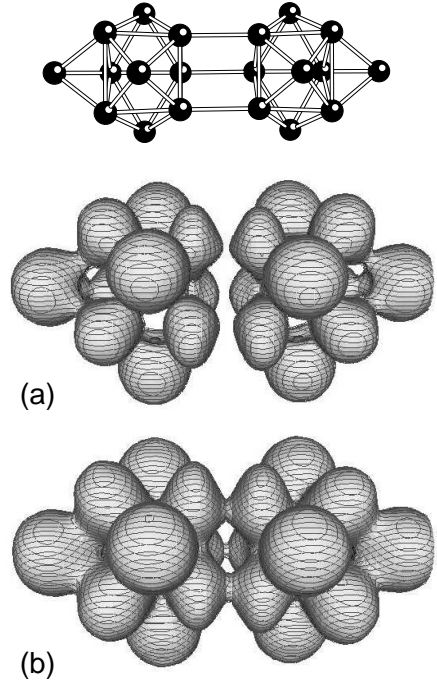}}
\caption{\label{fig2} Isosurfaces of the electron
localization function $\chi_{\rm ELF}$ for a low-lying prolate form
of Sn$_{20}$ containing two TTP units;  
(a) $\chi_{{\rm ELF}}=0.7$, (b) $\chi _{{\rm ELF}}=0.55$.\cite{explan}}
\end{figure}               

It is of considerable interest to investigate the nature of bonding in
the Sn$_{20}$ cluster.  It may be recalled that at room temperature
bulk Sn exists in a metallic phase (white tin), while in our previous
work \cite{sn10} we have shown Sn$_{10}$ to be a covalently bonded
cluster.  While we therefore expect the bonding within a TTP unit in
Sn$_{20}$ to be covalent, it is interesting to examine the bonding
between the TTP units.  In Fig.\ \ref{fig2} we show isosurfaces of the
ELF taken at two different values of $\chi _{{\rm ELF}}$ for a
typical prolate structure. Recall that a higher value of the ELF,
approaching unity, indicates a localized distribution of valence
electrons.  We note that the isosurface for the higher value $\chi
_{{\rm ELF}}=0.7$, shown in Fig.\ \ref{fig2}(a), connects the Sn ions
within each TTP unit along the line joining them, implying that the
ions within each TTP unit are covalently bonded.  However, this
higher-value isosurface of the ELF does not connect the two different
TTP units.  The bonds between the two TTP units are only observed at a
rather lower ELF value $\chi _{{\rm ELF}}\approx 0.55$, shown in
Fig.~\ref{fig2}(b).  Evidently the two TTP units are weakly bonded to
each other compared to the bonding within a TTP unit.  This leads to a
number of low-lying isomers with nearly degenerate energies,
of which some are shown
in Fig.\ \ref{fig1}(a), and which as noted above differ in the relative
orientation of the TTP units.

\begin{figure}
\epsfxsize=7.5cm
\centerline{\epsfbox{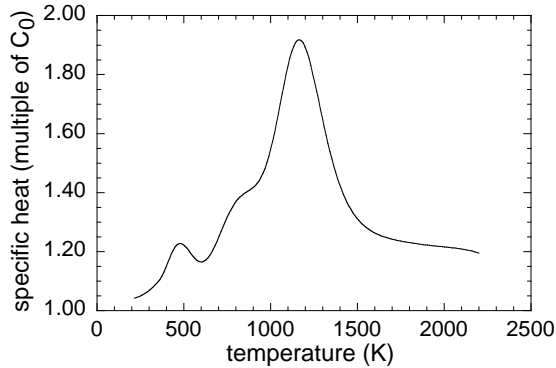}}
\caption{Canonical specific heat of Sn$_{20}$ cluster.  
$C_0=(3N-9/2)k_B$ is the zero-temperature classical limit of the rotational 
plus vibrational canonical specific 
heat.\protect\cite{nathermo}\label{fig3}}
\end{figure}               

Turning to the canonical ionic specific heat (shown in
Fig.~\ref{fig3}), we note some interesting features.  First, the main
peak in the specific heat, often identified with the ``melting'' of the
cluster, occurs around 1200~K, much higher than the bulk melting point
(505~K).  Second, there is also a small peak around 500~K and a
shoulder around 800~K. These features can be explained by examining
the microscopic processes as a function of temperature.  At low
temperatures (below 400~K), the only motion observed is a change in
the relative orientation of the two TTP units, which spans the class
of low-lying isomers [Fig.~\ref{fig1}(a)] having two distinct TTP
units.

However, at around 500~K, the atoms within each TTP unit start to
rearrange themselves in such a way that the TTP structure remains
intact, but undergoes a reshuffling of at least three atoms in every
rearrangement, including an interchange of capping atoms with atoms of
the trigonal prism core.  This process is similar to the one observed
in Sn$_{10}$.\cite {sn10} Since this rearrangement does not distort
the TTP unit, the shape of the cluster remains prolate. The
distortion of the TTP units themselves begins around 800~K. Starting
at this temperature, we observe that at least one of the TTP units may
be seriously distorted at various points in the dynamics, to the
extent of losing its identity.  The onset of these processes at 500~K
and 800~K is responsible for the features seen in the canonical
specific-heat curve.

\begin{figure}
\epsfxsize=7.5cm
\centerline{\epsfbox{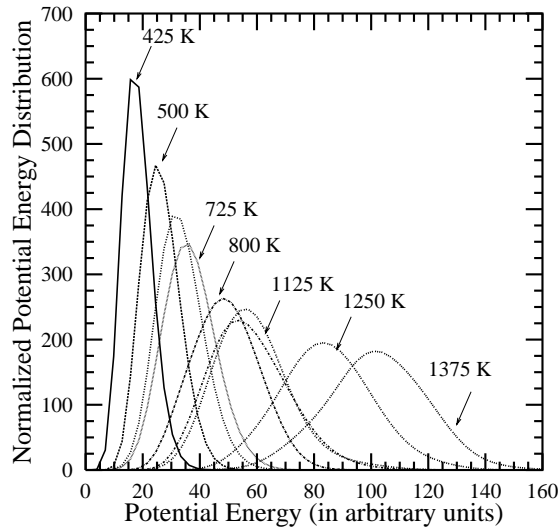}}
\caption{\label{fig4}Normalized potential-energy distributions for
400~K to 1375~K. The distributions are obtained from the actual data
of an isokinetic sampling run at the kinetic temperature shown, but
have been smoothed to suppress fluctuations and make the figure
clearer.}
\end{figure}               

This is also reflected in the potential-energy distributions (PED) for
these temperatures.  In Fig.\ \ref{fig4},
we show normalized PED for a series
of temperatures starting at 400~K and going up to about 1300~K.
Let us recall that the PED $p(E,T)$ at temperature $T$ follows the
Gibbs distribution $p(E,T)=\Omega (E)\exp (-E/k_{B}T)/Z(T)$, where
$Z(T)$ is the canonical partition function and $\Omega (E)$ is the
density of accessible states.  An examination of Fig.\ \ref{fig4}
reveals a slight relative broadening of the PED around 800~K, indicating
that other modes of excitation are also available at this temperature,
apart from the relative orientation of the TTP units and the internal
rearrangements within the TTP units that take place at lower
temperatures.  This broadening is reflected in the ionic specific heat
as a shoulder around 800~K.

As the temperature rises above 800~K, the relative motion between the
two TTP units becomes more vigorous.  The TTP units, apart from
undergoing internal distortion, also start to come sufficiently close
to one another as to initiate the exchange of atoms.  This process
ultimately leads at about 1200~K to the collapse of the prolate
structure toward a more compact structure such as those shown in
Fig.~\ref{fig1}(b).  In fact, in the temperature range 1200 to 1300~K,
we occasionally observe that the collapsed structure reforms into a
prolate structure.  Thus the time scale at these temperatures for
``hopping'' between a prolate and a more compact structure is of the
same order of magnitude as that of our sampling runs, namely, about
40~ps.

To summarize, there are three main modes of ionic motion observed
during our simulations up to 1200~K: (i) changes in the relative
orientation of two Sn$_{10}$ units, (ii) an internal rearrangement of
ions within each TTP unit, observed after 500~K, and (iii) the
distortion of at least one TTP unit accompanied by the interchange of
atoms between the two TTP units, observed after 800~K, leading
ultimately to the fusion of the TTP units and a shape change at
1200~K.

\begin{figure}
\epsfxsize=7.5cm
\centerline{\epsfbox{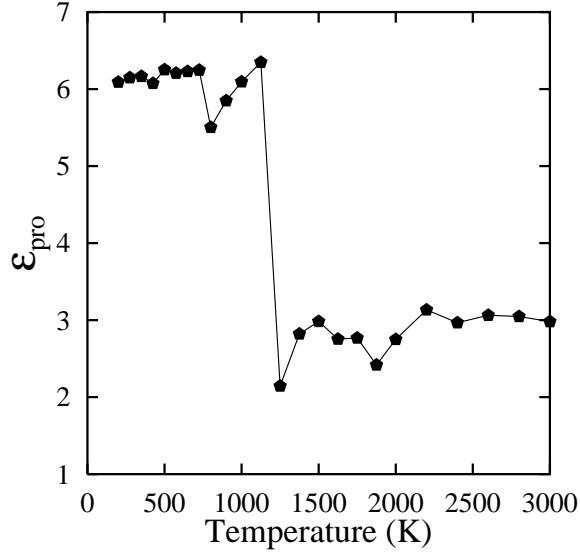}}
\caption{\label{fig5}  The coefficient $\epsilon_{\mathrm{pro}}$ 
[see Eq.~(\protect\ref{eqn:epspro})] describing the degree of prolate 
deformation of the cluster as a function of temperature. Note the 
significant decrease around 1250~K.}
\end{figure}               

Since it is the collapse of a prolate shape toward a more spherical
shape that causes a change in the diffusion coefficient in an ion
mobility experiment,\cite {Alex} it is interesting to investigate
these shape changes more explicitly.  In Fig.\ \ref{fig5}, we show the
thermally averaged value of the deformation coefficient
$\epsilon_{\mathrm{pro}}$, calculated for temperatures ranging from
200 to 3000~K, and averaged over 35~ps for each temperature.  For a
given ionic configuration,
$\epsilon_{\mathrm{pro}}$ is defined as
\begin{equation}
\epsilon_{\mathrm{pro}} = \frac{2Q_{1}}{Q_2+Q_3},
\label{eqn:epspro}
\end{equation}
where
$Q_1 \geq Q_2 \geq Q_3$ are the eigenvalues, in descending 
order, of the quadrupole tensor
\begin {equation}
	Q_{ij} =  {\sum_{I}  R_{Ii}\,R_{Ij} }.
\end {equation}
Here $i$ and $j$ run from 1 to 3, $I$ runs over the number of ions,
and $R_{Ii}$ is the $i$th coordinate of ion $I$ relative to the
cluster center of mass.  We note that $\epsilon_{\mathrm{pro}}$ will
be unity for a spherical cluster and greater than unity for prolate
deformations.  At very low temperatures, we see that
$\epsilon_{\mathrm{pro}}
\approx 6$, corresponding to the ground-state structure in
Fig.\ \ref{fig1}(a), which can be regarded roughly as having a
uniaxial prolate deformation with aspect ratio
$\sqrt{\epsilon_{\mathrm{pro}}} \approx 2.5$.  This value of
$\epsilon_{\mathrm{pro}}$ is maintained until about 800~K, where
fluctuations are observed, corresponding to the distortion of one or
both TTP units, as described above.  However, the cluster still
maintains an overall prolate shape until about 1250~K, when there is a
very sharp drop in $\epsilon_{\mathrm{pro}}$ toward a value of about 2
to 3, which is maintained up to temperatures of 3000~K or more.  This
drop corresponds to the fusion of the two TTP units to give the more
spherical shapes shown in Fig.\ \ref{fig1}(b).

\begin{figure}
\epsfxsize=7.5cm
\centerline{\epsfbox{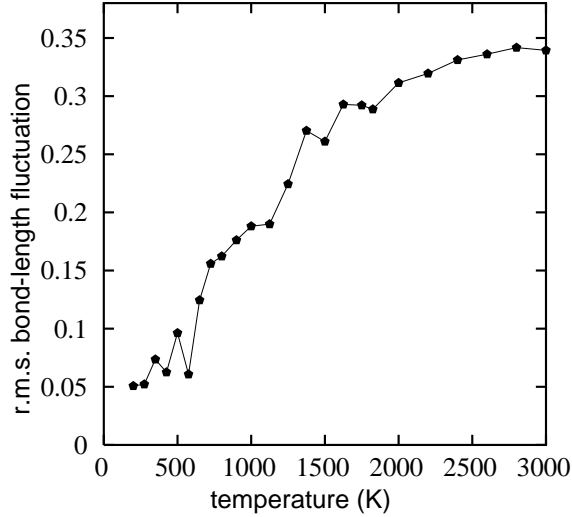}}
\caption{\label{fig6} The root-mean-square bond-length fluctuation
$\delta_{\mathrm{rms}}$ [Eq.~(\protect\ref{eqn:deltarms})] of Sn$_{20}$.}
\end{figure}

In discussions of cluster ``melting'' phenomena, it is often useful
to consider the root-mean-square bond-length fluctuation $\delta _{{\rm 
rms}}$, defind as
\begin{equation}
\delta _{{\rm rms}}=\frac{2}{N(N-1)}\sum_{I>J}\frac{(\langle
R_{IJ}^{2}\rangle _{t}-\langle R_{IJ}\rangle _{t}^{2})^{1/2}}{\langle
R_{IJ}\rangle _{t}},  \label{eqn:deltarms}
\end{equation}
where $N$ is the number of ions in the system, $R_{IJ}$ is the
distance between ions $I$ and $J$, and $\langle \cdots \rangle _{t}$
denotes a time average over the entire trajectory.  The calculated
$\delta_{\mathrm{rms}}$ is shown in Fig.~\ref{fig6}.  It can be
observed that the traditional Lindemann criterion of
$\delta_{\mathrm{rms}}=0.1$ signaling bulk melting is reached quite
early, around 650~K. This happens because $\delta_{\mathrm{rms}}$ is
sensitive to the permutational rearrangement of ions within a TTP unit
that occurs at these temperatures.  This highlights the difficulty of
giving a precise definition of ``melting'' in a small cluster.  In the
temperature range 400 to 1300~K, Sn$_{20}$ undergoes a series of
complex rearrangement processes.  In some sense, an individual TTP
unit could be said to have melted already at 500~K, since after a
sufficiently long time the reshuffling motion found at this
temperature allows ions to permute throughout a TTP unit.  After about
800~K, when movement of ions between the two TTP units becomes
possible, ions can permute throughout the whole cluster.  However, the
dominant peak in the specific heat and the collapse of the overall prolate
shape of the cluster do not occur until 1200~K. Note that a small step
in $\delta_{\mathrm{rms}}$ can also be observed near 1250~K, corresponding
to this collapse.  In this final, collapsed state, for temperatures of 
1300~K and upwards, the ionic motion is diffusive and the cluster has
a highly liquidlike character.

\begin{figure}
\epsfxsize=7.5cm
\centerline{\epsfbox{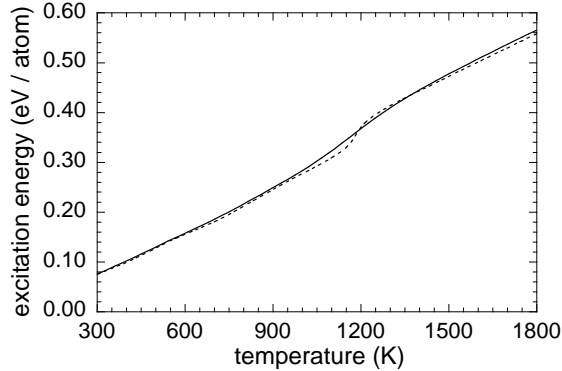}}
\caption{The canonical (solid line) and microcanonical (dashed) caloric 
curves for Sn$_{20}$.  In the microcanonical case, the temperature is defined as
$T(E)=[\partial S(E)/\partial E]^{-1}$ for excitation energy $E$.  In
the canonical case, the temperature is the heat-bath temperature, and
the excitation energy is the thermally averaged excitation 
energy.\label{fig7}}
\end{figure}               

\begin{figure}
\epsfxsize=7.5cm
\centerline{\epsfbox{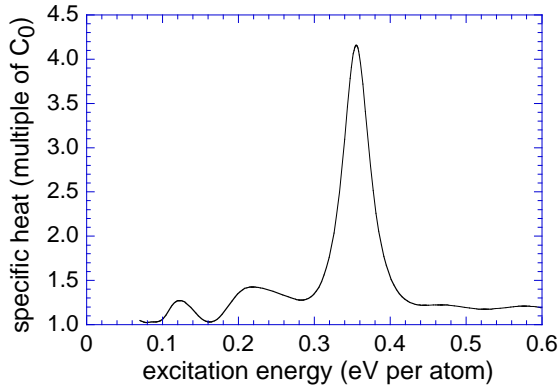}}
\caption{The microcanonical specific heat for Sn$_{20}$.  For $C_{0}$,
see caption to Fig.~\protect\ref{fig3}.\label{fig8}}
\end{figure}               

The canonical and microcanonical caloric curves are shown in
Fig.~\ref{fig7}.  The canonical curve shows a mild kink extending
across the temperature range 800 to 1400~K, corresponding to the peak
in the canonical specific heat in Fig.\ \ref{fig3}.  The
microcanonical curve has a somewhat more distinct step around 1250~K,
but the curve does not develop an S bend, as is theoretically possible
in the microcanonical ensemble.\cite{hill} No S bend was observed
either for Sn$_{10}$ in our previous work using the same LDA
model.\cite{sn10} As is well known, S bends for the solidlike to
liquidlike transition in clusters typically occur only for special
sizes, such as the complete Mackay icosahedra for the Lennard-Jones
potential.\cite{MH} Corresponding to the absence of an S bend in the
caloric curve, the microcanonical specific heat $C(E)$, shown in
Fig.~\ref{fig8}, is everywhere finite and positive in the energy range
studied.  The curve $C(E)$ shows the same basic features as the
canonical curve $C(T)$ in Fig.~\ref{fig3}, but with higher resolution. 
In particular, the shoulder at 800~K and the main peak at 1200~K in
$C(T)$ become clearly resolved in $C(E)$ as a small peak at an
excitation energy $E=0.22$~eV per atom and a dominant peak at 0.36~eV per
atom, respectively.  This allows an extraction of the effective
``latent heat'' $L$ for the collapse of the prolate form, $L\approx
0.042$~eV per atom, taken to be the area under the main peak in $C(E)$
and above a line joining the points on the $C(E)$ curve at $E=0.3$ and
$E=0.42$~eV per atom.

\begin{figure}
\epsfxsize=7.5cm
\centerline{\epsfbox{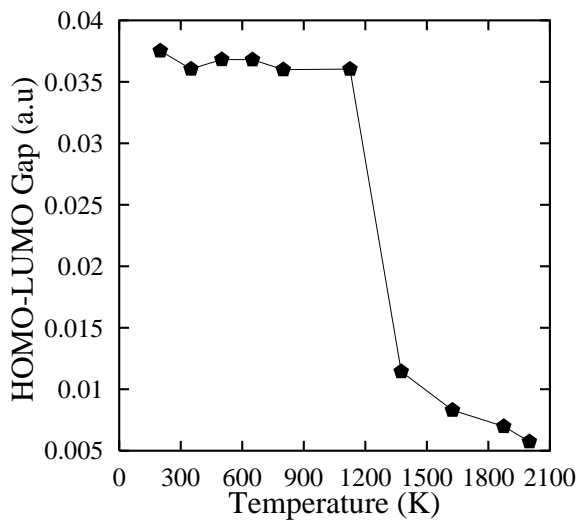}}
\caption{\label{fig9}  The HOMO-LUMO gap, averaged over 7.5~ps, as a 
function of temperature.}
\end{figure}               

Interestingly, the collapse of the prolate form around 1250~K is
accompanied by a rather drastic reduction in the HOMO-LUMO gap.  In
Fig.\ \ref{fig9}, we show the thermal average (taken over 7.5 ps) of
the HOMO-LUMO gap as a function of temperature.  The gap, which is fairly
constant at about 0.036~a.u.\ up to 1200~K, drops sharply by almost a
factor of four around 1250~K, and continues to decrease as the
temperature rises further.  At first sight, it is tempting to
interpret this reduction in the gap as related to the change in the
nature of bonding from covalent to delocalized metallic.

\begin{figure}
\epsfxsize=7.5cm
\centerline{\epsfbox{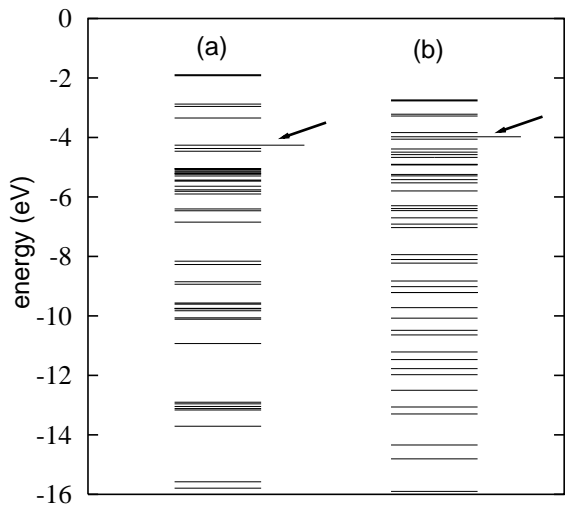}}
\caption{\label{fig10} The eigenvalue spectrum for (a) the ground state 
and (b) a nonprolate structure occurring at high
temperature. The arrow indicates 40th state.}
\end{figure}               

\begin{figure}
\epsfxsize=7.5cm
\centerline{\epsfbox{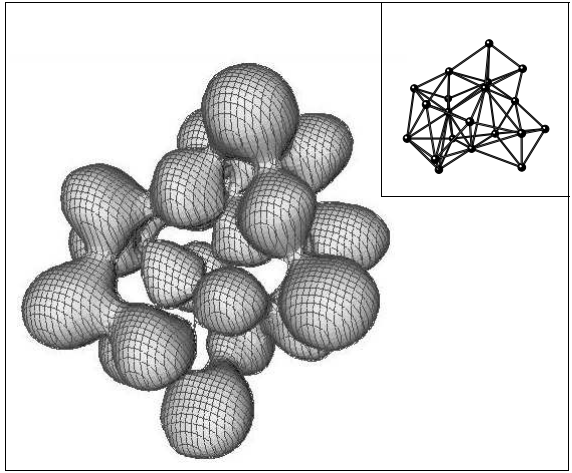}}
\caption{\label{fig11} Isosurface of the electron
localization function $\chi_{\rm ELF}$ for a randomly chosen 
high-temperature structure of Sn$_{20}$. $\chi _{{\rm ELF}}=0.7$ and 
the structure is shown in the inset.}
\end{figure}

To investigate this point in more detail, we have examined the
eigenvalue spectrum of a typical prolate and nonprolate structure,
along with an analysis of the number of nearest neighbors of these
structures.  Although the nonprolate structure is more compact, the
average number of nearest neighbors in both the prolate and nonprolate
structures is almost the same.  However, in the case of nonprolate
structures, the bond angles are not optimal, resulting in strained
bonds.  These strained bonds can be considered to introduce disorder,
which spreads out the energy levels by lifting degeneracies and
splitting groupings of energy levels.  This is evident from the
Kohn-Sham eigenvalue spectrum [shown in Fig.\ \ref{fig10}] of a
prolate and nonprolate structure.  Although the total occupied
bandwidth is almost the same, the eigenvalue spectrum of the prolate
structure shows distinct bands, in contrast to the spread-out
eigenvalue spectrum of the nonprolate structure, leading to the
observed substantial reduction in the HOMO-LUMO gap.  To demonstrate
that this reduction in the HOMO-LUMO gap is due to the disorder in the
system and not to a change in the nature of the bonding, we have also
examined the ELF of some randomly chosen high-temperature nonprolate
structures.  An isosurface of the ELF for one of these structures is
shown in Fig.\ \ref{fig11} at a value $\chi _{{\rm ELF}}=0.7$.  Quite
clearly, there is a significant localization of electron density between
nearest neighbors, and the bonding is covalent.

Finally, it is interesting to note that, although in Sn$_{20}$ the
dominant peak in the specific heat occurs at a much higher temperature
(1250~K) than the bulk melting point (505~K), the peak for Sn$_{10}$
occurs at a still higher temperature (2300~K).\cite{sn10} To
understand this, we note some features of the two clusters.  Sn$_{10}$
contains one TTP unit, whereas Sn$_{20}$ contains two.  Although the
nature of the bonding within each TTP unit is covalent, in Sn$_{20}$
the two TTP units are weakly bonded to each other.  When ions acquire
sufficient kinetic energy, they can cross from one TTP unit to the
other, and the two TTP units in Sn$_{20}$ lose their separate
identity.  That is, the presence of another TTP unit provides an extra
channel for ions to move.  In Sn$_{10}$, the main peak in the specific
heat is associated with the breaking of covalent bonds within the TTP
unit.

In future work, it will be interesting to investigate the
transition from a prolate ground state to a nonprolate ground state as
a function of cluster size.

\section {Conclusions\label{sec:conclusions}}

The present {\em ab initio} simulations on Sn$_{20}$, along with
earlier work on Sn$_{10}$,\cite{sn10} provide insights and an
explanation of the higher-than-bulk ``melting point'' observed
experimentally.\cite{Alex} There are two basic reasons for this
behavior: (i) the covalent nature of the Sn-Sn bond in the cluster,
compared to the metallic bonding of bulk Sn at room temperature, and
(ii) the special stability of the TTP unit.  We also note that the
``melting'' of a cluster as detected in an ion mobility experiment is
specifically associated with the collapse of the prolate cluster
shape, which requires a collapse of the TTP units themselves. 
Although many rearrangement processes can occur at lower temperatures,
the dominant feature of the specific heat (or caloric curve) for both
Sn$_{10}$ and Sn$_{20}$ is associated with the break up of the TTP
structure.  In the cluster containing one TTP unit, Sn$_{10}$, the TTP unit
breaks up at the higher temperature, since the only processes that
destroy the TTP unit are the ones that break the bonds.  As the size
increases, for example Sn$_{20}$, containing two TTP units, the
collapse of the two TTP units leading to a compact cluster is aided by
a relative motion between them, and occurs at a much lower
temperature than the break up of the TTP unit for Sn$_{10}$.

In a general way, in a prolate cluster there will be a competition
between the covalent bonding, which tends to favor stacks of TTP
units, and the surface energy, which tends to favor more spherical
structures.  Our calculations reveal that for Sn$_{20}$ the surface
energy does not become the dominant factor until temperatures of about
1200~K. It is reasonable to expect that as the cluster size increases,
the temperature for the collapse of the prolate structure will reduce
further, until eventually the ground state structure itself is no longer
prolate, as observed experimentally.

\acknowledgments

We gratefully acknowledge the support of the Indo-French
Center for the Promotion of Advanced Research under
Project No.\ 1901-1.


\begin{thebibliography}{}
\bibitem{explan} We have examined the ELF for all the structures in
Fig.\ \ref{fig1}(a) and have found them to be essentially similar.
\end{thebibliography}
\end{document}